# Red Flags and Cherry Picking: Reading The Scientific Blackpill Wiki

Celia Chen, Alex Leitch, Scotty Beland, Ingo Burghardt, William Conway, Rajesh Kumar Gnanasekaran, Marilyn Harbert, Emily Klein and Jennifer Golbeck

## ABSTRACT

Incels are an online community of men who share a belief in extreme misogyny, the glorification of violence, and biological essentialism. They refer to their core ideology as "The Blackpill", a belief that physical attraction is the only path to romantic success and that women are only attracted to one very specific, hypermasculine archetype. This is not only a belief system; incels believe their ideology grounded in hard science. The research that incels use as evidence of their belief system is collected in an extensive online document, the Scientific Blackpill wiki page. In this research, we analyze the claims made on the wiki against the research cited to assess how the wiki authors are using or misusing science in support of their ideology. We find that the page largely cites legitimate science and describes it partly or mostly accurately. However, in discussing it, the results are often overgeneralized, stripped of context, or otherwise distorted to support the pre-existing incel viewpoint. This echoes previous findings about motivated reasoning and borrowing scientific legitimacy in other misinformation and conspiracy-minded ideologies. We discuss the implications this has for understanding online radicalization and information quality.

## CCS CONCEPTS

• Social and professional topics User characteristics

## 1 Introduction

The incel community is an online, violent, misogynistic hate movement comprising straight men who are unable to connect romantically with women and who blame society at large and women specifically for their failures. Members of the community have been responsible for mass shootings and other acts of public violence as part of the community's ideology.

Becoming radicalized into inceldom is referred to as "taking the blackpill" [1]. This means adopting a set of beliefs in which genetics and appearance are the sole determinant of romantic success, and specifically that women are only interested in hypermasculine, white men with features that are typically Aryan (though the community generally does not use that term). The blackpill represents a fatalistic worldview where personal agency is rejected and systemic problems are deemed unsolvable through individual action [2].

This paper investigates the Scientific Blackpill page on an incel-community maintained wiki. This 117,000-word wiki page presents a detailed argument justifying the blackpill worldview with scientific research. Organized into 18 sections with 219 subsections, the page cites 578 articles, news stories, technical reports, and blog posts in its exploration of categories like women's level of attraction to men based on facial structure, height, and race; preferences regarding penis size; the role of mental health in romantic success; women's commitment in relationships; and rape myths.

The blackpill worldview is prima facie incorrect; men who are short, bald, poor, who lack traditional hypermasculine facial features, who are neurodivergent, or who are non-white regularly find enthusiastic sexual partners and meaningful romantic connection. However, the Scientific Blackpill wiki's compilation of citations to peer-reviewed research creates a veneer of scientific legitimacy that may make these extreme claims more persuasive to vulnerable individuals seeking explanations for their personal romantic difficulties.

Despite growing research on incel communities and radicalization processes [1][2][3], no systemic analysis has examined how the Scientific Blackpill distorts research to support its extremist ideology.

This study addresses the following research questions:

- How does the Scientific Blackpill page attempt to construct a veneer of scientific legitimacy for incel ideology through its selection, citation, and interpretation of research?
- How are scientific concepts misused, decontextualized, or selectively invoked to reinforce the blackpill worldview?
- How does the construction and use of this page support the broader radicalization process within incel communities?

## 2 Related Work

### 2.1 Incel Communities and Radicalization

Incels (involuntary celibates) are an extremist online community unified by their inability to establish romantic relationships with women and characterized by misogyny, racism, and glorification of violence [3]. The community congregates almost entirely online across forums like incels.is, creating echo chambers where extreme beliefs are normalized and reinforced. Members overwhelmingly voice support for acts of misogynistic violence, leading scholars to consider incel violence as a new variety of terrorism with a specific hate crime dimension [5].

Research on incel radicalization has identified several key processes. Golbeck et al. [1] conducted a thematic analysis of incels' self-described radicalization narratives, finding a four-stage chronological process:
(1) pre-radicalization, marked by appearance concerns, social isolation, and psychological struggles;
(2) searching for blame, where incels identify society, women, genetics, or autism as culprits for their difficulties;
(3) radicalization, often catalyzed by YouTube content and producing a "breakthrough realization"; and
(4) post-radicalization, characterized by feelings of liberation, clarity, and community belonging.

These stages align closely with established models of extremist radicalization in other contexts [14], suggesting incels follow predictable psychological pathways toward extremism.

The "blackpill pipeline" is a progression through increasingly radical ideological positions [2]. New members enter through the "redpill" (recognition of perceived injustice), progress through the "racepill" (acceptance of racial hierarchies), and culminate in the "blackpill" (fatalistic belief that systemic problems cannot be solved, leading to advocacy for violence). This pipeline functions as a filtering mechanism where each stage removes agency and concentrates the most susceptible members toward extreme positions.

Previous work in this area [3] developed a lexicon of incel-specific terminology, revealing how specialized language serves to mark group membership, obscure meaning from outsiders, and provide a "combinatorial toolkit" for expressing increasingly extreme beliefs. The prevalence of dehumanizing (71.8%), racist (26.6%) and misogynistic (26.6%) terminology reflects the community's core values and provides markers for tracking radicalization over time.

## 2.2 Science Misuse and Motivated Reasoning

Motivated reasoning, or the tendency for people to arrive at conclusions they want to reach rather than conclusions warranted by evidence, provides a psychological foundation for understanding how scientific evidence gets distorted. When motivated to reach particular conclusions, individuals rely on biased cognitive strategies for accessing, constructing, and evaluating beliefs, enhancing use of those strategies most likely to yield desired conclusions while their ability to do so remains constrained by their need to construct seemingly reasonable justifications [6]. This motivated cognition leads people to reject scientific findings that threaten core beliefs or worldviews [7]. Research on scientific denialism has identified five characteristic tactics: invoking conspiracy theories about scientific consensus, deploying fake experts, selectively cherry-picking evidence, creating impossible expectations for certainty, and employing misrepresentation through logical fallacies [8].

When individuals become ideologically committed to particular positions, they engage in biased information-processing strategies that reliably guide them toward beliefs congruent with their identity-defining groups [9]. This identity-protective cognition leads individuals to selectively search for confirming evidence, critically scrutinize disconfirming information while accepting confirming information at face value, and interpret ambiguous evidence in ways that support predetermined conclusions. The process is amplified in online echo chambers where confirmation bias receives social reinforcement.

Historical examples illustrate how scientific language can be weaponized to justify discrimination. Nineteenth-century craniometry claimed to provide objective evidence for racial hierarchies; social Darwinism misapplied evolutionary theory to justify inequality; and contemporary movements from climate change denial to anti-vaccination campaigns selectively cite research while ignoring contradictory evidence [10]. In each case, advocates create an appearance of scientific legitimacy through selective citation while distorting the research findings.

Research on evolutionary psychology, a field heavily cited in the Scientific Blackpill wiki, has documented how legitimate findings are often misrepresented in popular discourse to support gender essentialism and biological determinism. The gap between research findings that acknowledge within-group variations, cultural influences, and developmental plasticity and public perception emphasizing deterministic, between-group differences creates opportunities for ideological exploitation.

## 2.3 The Scientific Blackpill as a Unique Case

While science misuse has been studied in contexts like vaccine hesitancy, climate denial, and creationism, the Scientific Blackpill wiki represents a case that merits its own analysis. The wiki's extensive compilation of academic citations across hundreds of subcategories suggests coordinated curation rather than the casual reference collection typical of most online extremist content. The nature and quality of these sources, and how they are deployed, remains an open empirical question central to understanding the wiki's persuasive mechanisms.

The wiki's role within the broader incel radicalization ecosystem distinguishes it from other forms of scientific misappropriation. While the blackpill pipeline described by [2] contextualizes how members progress toward fatalistic extremism, the specific role of the Scientific Blackpill wiki within this process has not yet been examined. The connection between incel ideology and real-world violence, including multiple mass casualty attacks, necessitates analysis of the mechanisms that legitimize and promote such extremism [5]. Yet no analysis has documented how the Scientific Blackpill distorts research, what patterns characterize its misrepresentation strategies, or how these distortions function within the radicalization process. This study addresses that gap through content analysis of the wiki's citation practices and source use.

## 3  Dataset

The Scientific Blackpill page of the Incel.wiki website is 116,949 words long - roughly 290 pages if printed. The page describes itself as follows:

> The Scientific Blackpill is about understanding the nature of human social and sexual behavior with a particular focus on evolutionary psychological perspectives.
>
> As per the blackpill, this compilation emphasizes the role of systemic and genetic factors and traits in men's dating issues (rather than personal ones). These include innate behavior and preferences, physical attractiveness, facial bone structure, stature, muscularity, body frame size, race, personality, local sex ratios, intelligence, ability, health, mental health, social and economic status, as well as female coyness, sneakiness and nastiness.
>
> This page maintains a neutral tone and conveys the scientific findings without judgment. However, sections demarcated as "discussion" may occasionally contain unsourced speculation or writing from a more non-neutral perspective.

The page is divided into 18 sections with 219 subsections. Each subsection has a topic or theme, e.g. "Male gang members have dramatically more female sexual partners". Each subsection comprises a summary of the research related to the topic. This is usually followed by a Discussion which includes commentary and analysis. The subsection may also have a set of Quotes pulled from the supporting research. Every subsection has a References list with citations to research that backs the idea of the subsection. The average subsection is around 500 words long, though there is a lot of variation. Most subsections cite to one or two outside sources, though some sections have more than 10.

The top level sections, breadth, depth, and topics of each are as follows:

1. **Personality** - 21 subsections, 78 citations
   Topics are women's attraction to traditionally "dark" personality traits like narcissism and psychopathy, the sexual appeal of criminals, but many of sub-sections are about women's interest in rape porn.
   E.g. "1.5 50% of female porn viewers admitted to watching porn involving extreme violence against women"

2. **Mental** - 10 subsections, 23 citations
   Focuses on mental health and relationship success with an emphasis on autism.
   E.g. ".2 44.6% of high functioning adult autistic men remain virgins, despite high sex/relationship drive"

3. **Race** - 17 subsections, 40 citations
   Focuses on challenges various races face with romantic success, focused on the success of whites and struggle of Asians.
   E.g. "3.3 Women are more racist than men in speed dating, and find Asian men least physically attractive"

4. **Looks** (Life) - 10 subsections, 34 citations
   Focuses on the way attractiveness impacts life.
   E.g. " 4.5 Parents treat attractive children better than ugly children"

5. **Looks** (Love) 18 subsections, 27 citations
   Focuses on the way attractiveness impacts romantic success,
   E.g. " 5.5 Only a man's looks and race matter in online dating - his personality does not"

6. **Face** - 14 subsections, 26 citations
   Focuses on facial features and how they relate to romantic and general life success.
   E.g. " 6.4 Teenage boys with 'dominant' facial features have sex earlier"

7. **Money** - 8 subsections, 14 citations
   Focuses on the role a man's income has in relationship success.
   E.g. " 7.3 Women orgasm more when having sex with rich men"

8. **Height** - 11 subsections, 21 citations
   Focuses on the role a man's height plays in his romantic and life success.
   E.g. " 8.9 Taller men report more satisfaction in their romantic relationships than shorter men"

9. **Body** - 9 subsections, 21 citations
   Focuses on strength, weight, disability, and physicality.
   E.g. "9.7 Among university students, only physical dominance over other men predicted mating success"

10. **Penis** - 4 subsections, 4 citations
    Focuses on women's preference for penis size.
    E.g. "10.4 90% of women agree that penis girth is more important than length for their sexual satisfaction"

11. **Voice** - 4 subsections, 7 citations
    Focuses on how voice pitch impacts romantic and life success.
    E.g. "11.2 Social dialect and men's voice pitch influence women's mate"

12. **Age** - 9 subsections, 21 citations
    Focuses largely on men being attracted to younger women, especially underage girls.

E.g. "12.1 It is normal for healthy men to find pubescent & prepubescent females sexually arousing"

13. **Hypergamy** - 18 subsections, 60 citations
Focuses on women's interest in marrying men of higher status, especially the imbalance between men's interest in women and women's interest in men.
E.g. "13.5 A survey found a dramatically higher median sex partner count for young women than young men"

14. **Cucks** - 5 subsections, 18 citations
Focuses on women's loss of interest in relationship, including cheating and divorce.
E.g. "14.3 The more women love their husbands, the less likely they are to initiate sex"

15. **Sluts** - 13 subsections, 25 citations
Focuses on women's sexual activity.
E.g. "15.9 Women's reported sex partner count dramatically increases when hooked up to a polygraph"

16. **MeToo** - 12 subsections, 26 citations
Focuses on the idea that women exaggerate and fabricate accusations of sexual harassment and assault, that this is burdensome to men, and that men are more often victims of violence in relationships.
E.g. "16.5 Men & especially ugly men are considered inherently 'creepier' than women"

17. **Health** - 15 subsections, 20 citations
Focuses on the relationship between romantic / sexual connection and overall physical, social, and mental health.
E.g. "17.4 Penile–vaginal intercourse is associated with health, but masturbation and anal sex are not"

18. **ItsOver** - 21 subsections, 113 citations
Focuses on hopelessness attributable to blackpill ideas being true, along with suicide.
E.g. "18.20 Involuntarily celibates often were ostracized, bullied, and socially withdrawn during childhood"

## 4 Methods

### 4.1 Codebook Development

The codebook underwent iterative development to ensure reliability and comprehensiveness in capturing distortion patterns. Initial codebook construction (v1.0) began with the lead author independently coding the first major section of the wiki, documenting emergent patterns of misrepresentation and establishing preliminary categories. This initial coding included nine fields: wiki section/subsection location, paper title, article type, publication year, actual finding (limited to 2-3 sentences), wiki representation, distortion type (with six categories: Accurate, Cherry-picking, Overgeneralization, Context Stripping, Loaded Language, and Statistical Manipulation), red flags for study quality issues, and additional notes.

To refine the codebook, we conducted a pilot coding exercise with most of the full research team (n=8 coders). The senior researcher selected a deliberately brief wiki section on penis size preferences for this calibration exercise, chosen both for its brevity (four citations) and as an early assessment of coders' comfort engaging with the wiki's content. Team members independently coded this section and provided structured feedback on category definitions, boundary cases, and needed additions or clarifications.

The research team operated remotely coordinating via email listserv. Following the pilot coding exercise, team members shared their coded spreadsheets and engaged in an asynchronous discussion identifying challenges and proposing refinements. Key issues raised by multiple coders included the need to distinguish factual accuracy from interpretive distortion (leading to the separation of "Veracity" from "Distortion Type"), handling section titles that made stronger claims than the body text, addressing completely fabricated claims versus selective misrepresentation, and categorizing ad hominem attacks on researchers as different from other distortion types. Through this email-based deliberation, the team reached consensus on major codebook revisions (v2.0) including:

1. Adding a dedicated field for section titles to track instances where titles made more extreme claims than body content,
2. Separating factual accuracy ("Veracity" with three levels: Accurate, Partially True/Misleading, False/Deceptive) from interpretive distortion ("Distortion Type"),
3. Establishing hierarchical severity levels for study quality issues (Critical/Major/Moderate/Limitations),
4. Expanding distortion categories to include Fabrication (with three subtypes distinguishing invented statistics, contradicted sources, and false attribution), Correlation Misinterpretation, Irrelevant Information (capturing ad hominem attacks and credibility undermining), and False Balance,
5. Creating a field to document the wiki's self-acknowledged limitations (None / Minimal/ Substantial/ False Balance), and
6. Expanding guidance for the "Actual Finding" field to allow longer summaries for studies and encourage direct quotation of author caveats.

### 4.2 Data Coding

Our research team analyzed the Scientific Blackpill page by reading every subsection and external reference cited, and coding the attributes of the discussion according to the codebook.

Following codebook revision to v2.0, the lead author distributed the updated instrument concurrently with section assignments for the full wiki analysis. Team members proceeded directly to coding

Insert Your Title Here                                                                            WOODSTOCK'18, June, 2018, El Paso, Texas USAtheir assigned sections without re-coding the pilot material. No coders reported difficulties with the revised codebook during the month-long coding process. The lead author later identified the need to retroactively re-code both the initial test section (coded with v1.0) and the pilot section to ensure consistency across the full dataset.

## 5  Results

### 5.1    Scientific Content

Overall, authors on the Scientific Blackpill wiki were citing high-quality, peer-reviewed scientific journal articles, representing 68% of all sources. The presentation of this research was rarely presented incorrectly; it was presented accurately or partially misleadingly the vast majority of the time. Most of the distortions of the research findings derived from fitting the results to the incel narrative by stripping it of context, overgeneralizing, or describing it with loaded language.

 For example, consider section 14.2: Cucks: Women rapidly lose interest in sex once in a stable relationship or living with a man. This begins by discussing the work of Klusmann [11], a peer-reviewed journal article that examines how interest in sex relates to relationship duration across age groups. The results were mixed, finding that men's motivation for sex remains fairly constant over time while women's varies, but other variables did not show this difference. The article proposes an evolutionary psychology hypothesis to explain the finding, suggesting that women use sex to maintain a bond with their husbands in order to "procure male resources". While the article ignores biological factors that could impact women's sexual motivation over time like menopause, which affects all the women in their sample of couples aged 60+, and interpersonal that could affect sexual interest in the relationship like the impact of infidelity (over 40% of men in the oldest age group admit to having had an affair during the relationship compared to just 18% of women), tenderness (men report their interest in tenderness declines over time while women's remains the same, which could suggest women's interest in sex may decline as their other emotional needs are not being met), or the quality of sex they have with their partners, the paper presents the results of their large scale survey accurately and with statistical analyses that are sound.

The Scientific Blackpill wiki presents a fairly accurate 311 word summary of this article. It then presents a 1,010 word discussion that begins, "Contrary to common dating advice emphasizing soul matching and deep intimacy, the result suggests that, on average, a high degree of familiarity is actually detrimental to relationship stability because the woman gets bored, possibly due to the man opening up, revealing his flaws which is a low status signal." This sentence contains a number of incel themes - criticism of mainstream dating advice, rejection of the ability for men and women to emotionally connect, critique of women as "bored", the idea that if men are emotional they are "low status", and the belief that women are only interested in men who are "high status".

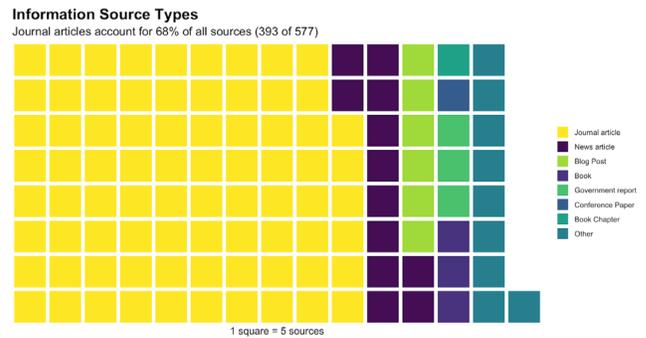

**Figure 1. Information source types cited on the wiki**

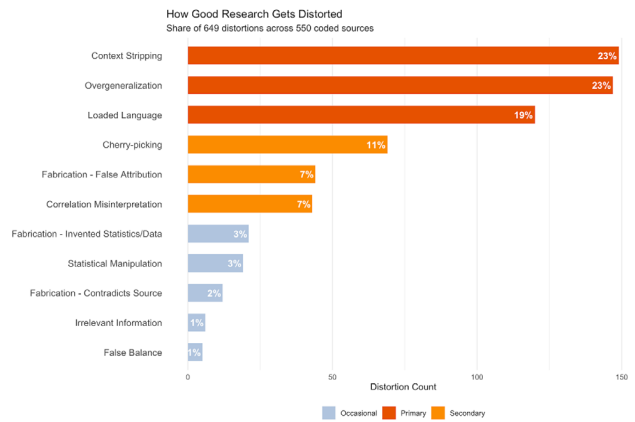

**Figure 2. Distortions of cited research**

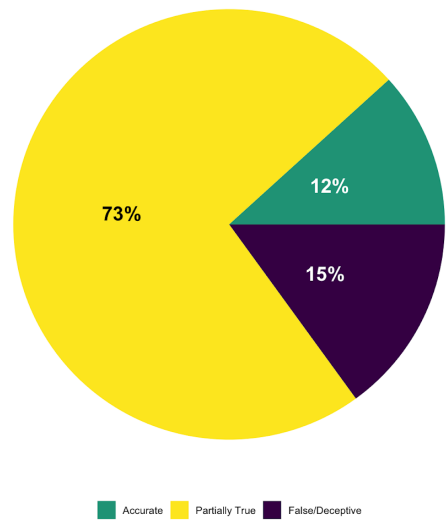

**Figure 3. Veracity of research presentation**

| Distortion Type | Partially True | Accurate | False/Deceptive |
|---|---|---|---|
| **Framing Distortions** | | | |
| Context Stripping | 115 | 32 | 11 |
| Overgeneralization | 130 | 11 | 8 |
| Loaded Language | 81 | 7 | 23 |
| Cherry-picking | 55 | 14 | 6 |
| Aggregation Distortion | 4 | 0 | 0 |
| Deterministic Framing | 9 | 0 | 0 |
| Essentialism | 1 | 0 | 0 |
| Exaggeration | 5 | 0 | 0 |
| False Balance | 2 | 4 | 0 |
| Mischaracterization | 6 | 1 | 1 |
| **Technical Errors** | | | |
| Correlation Misinterpretation | 37 | 0 | 5 |
| Statistical Manipulation | 9 | 0 | 2 |
| Irrelevant Information | 6 | 4 | 3 |
| **Fabrication** | | | |
| Fabrication - Contradicts Source | 3 | 0 | 10 |
| Fabrication - False Attribution | 22 | 1 | 20 |
| Fabrication - Invented Statistics/Data | 5 | 4 | 10 |
| Fabrication - Other | 3 | 1 | 2 |

Highlighted: >40% Partially True (yellow), >18% Accurate (green), >18% False/Deceptive (purple)

The writing in this wiki section is coded as Accurate because the results of the scientific papers are accurately described. The discussion around the science is speculative and reveals motivated reasoning, illustrating how the authors borrow scientific authority to support an incel narrative. This is typical of most discussions on the wiki.

An example of a Partially True presentation is in section 3.5 Race: Being Asian in the USA is a primary predictor of 'never being kissed'. This cites research [13] that shows, for a sample of first year undergraduates at a U.S. institution, 28% of Asian-American students reported never having been kissed vs. 7-11% for White, Black, and Latino students. While the research does show Asian-American students are more likely to have never kissed, it does not show that race is a "primary predictor". We coded this as "Statistical Manipulation" for misrepresenting the magnitude of the effect.

## 5.2 Page Development

The Scientific Blackpill page has had three main editors: BlackpillScience, Bibipi, and Altmark22. Together, they comprise 90% of the content added to the site and 95% of the number of edits made. Edits from non-logged in users accounted for an additional 8.8% of the content added. This overlaps with active periods from two of the more active users which could suggest those edits came from those users in logged out status. This period overlapped with linear growth in traffic to the wiki, which spiked aggressively between December 2023 and January 2024. This traffic bump is difficult to correctly interpret, as this period overlaps the documented growth in content scraping from companies producing data for large language models.

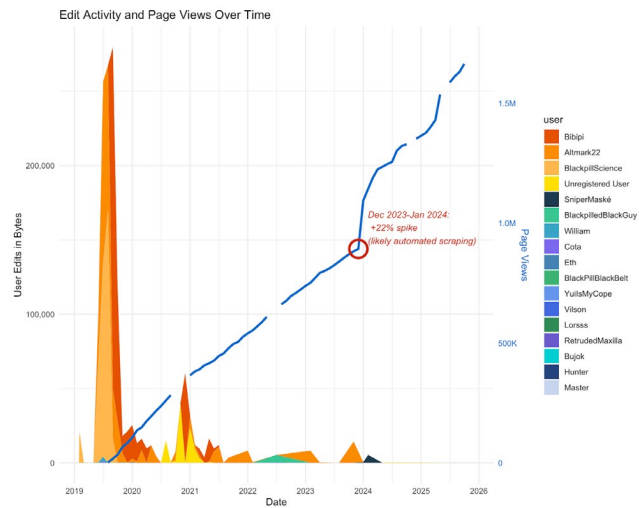

Figure 4. Edit activity and page views over time

## 6 Discussion

There are several ways to view these findings, but a key one is that the core edits to this wiki were made by three main accounts in and around the 2020 covid pandemic. These changes appear to support one another, with fascination attaching to specific sub-topics that are common between the accounts.

A notable trend in the wiki is the hard tension between an appearance of rationalism and a real drive to misrepresent the findings within the selected journal articles. This community places a heavy emphasis, even within the page itself, on the idea that women aren't rational, that rationality is itself a masculine value, and that women can be easily persuaded by "dark triad" personality traits, narcissism, and so on.

The data sources selected, while sometimes questionable, dated, or sexist themselves, nonetheless do typically represent scientifically grounded work. This is not how they are represented within the wiki.

In the wiki's actual writing, cherry-picking is deployed to support a worldview consistent with the one represented and continuously reinforced within various incel-focused social spaces online. This suggests that readings of this work are being driven by emotional patterns rather than by genuinely rational consideration of their findings. This has implications for the broader production of hierarchical authority through selective interpretation of data.

These tactics appear to be a type of wallpaper, protecting the editors engaging in such practice from considering deeper or more structural issues with their beliefs. This type of motivated reasoning and scientific misuse is common in other conspiratorial settings, with implications for interpretation of other misinformation collectives that adopt scientific or otherwise quantified rhetoric as legitimation for inaccurate beliefs, such as major-event denialist groups, including flat earthers, anti-vaxxers, and other categories of conspiracy theorist.

Here, the comfort of the numbers seems to act as a space for control over external phenomena, describing a psychological space where a facility with quantitative approaches can automatically stand in for "being rational," which is in turn reinforced through a directed editorial approach that prioritizes inclusion of quantitative works at the expense of their actual findings. This use of quantitative analysis to show achievement then circularly reinforces the hierarchical value of such approaches to prove masculinity.

Research from psychology, philosophy, and science communication have shown that individuals and groups often deploy the rhetorical style of science, with citations, quantitative language, and selective data interpretation, as a strategy for legitimizing inaccurate beliefs. Our analysis of the Scientific Blackpill aligns with those results. The patterns identified in the Scientific Blackpill suggest how online radicalization can be driven not only by exposure to extreme ideas but by encountering those ideas embedded within an apparently rigorous evidentiary framework. Given previous work that finds incels follow common radicalization processes, and that exposure to online content espousing the incel ideology is critical in the radicalization process [1], this kind of presentation of science-adjacent content allows vulnerable readers to rationalize increasingly extreme beliefs as "evidence-based." This dynamic shows how radicalization can proceed through cognitive co-option rather than persuasion; individuals adopt the movement's epistemic rules before adopting its most extreme ideological positions.

These findings also underscore broader concerns about information quality in digital environments where motivated communities can produce large, citation-dense resources that appear authoritative while being structurally misleading. The Scientific Blackpill demonstrates that misinformation is not simply false information presented as true, but the distortion of real information toward predetermined ideological ends. Our results echo findings about the misuse of science among climate deniers and covid deniers, integrating incel ideology into the broader context of ongoing scientific misinformation challenges.

## ACKNOWLEDGMENTS

This work was supported by the National Science Foundation under Grant No. <ANONYMIZED>.